\documentclass[11pt]{article}

\usepackage{amsfonts,amssymb,chet,dsfont,graphicx,mathrsfs,pgfplots,tikz,import,caption,subcaption}
\allowdisplaybreaks

\newcommand{\cs}{CE$\nu$NS}
\newcommand{\tabincell}[2]{\begin{tabular}{@{}#1@{}}#2\end{tabular}}

\title{Electroweak Constraints\\ from the COHERENT Experiment}

\author{Witold Skiba$^{\ast,}$\email{witold.skiba@yale.edu} and Qing Xia$^{\ast \dag,}$\email{qingxia@lbl.gov}}

\affiliation{
$^\ast$Department of Physics, Yale University, New Haven, CT 06520, USA \\
$^\dag$Lawrence Berkeley National Laboratory, Berkeley, CA 94720, USA }

\abstract{We compute bounds on coefficients of effective operators in the Standard Model that can be inferred from observations of neutrino scattering by the COHERENT experiment. While many operators are bound extremely well by past experiments the full future data set from COHERENT will provide modest improvements for some operators.}

 \date{October 2022} 

\begin{document}

\maketitle


\section{Introduction}\label{SecIntro}

Coherent neutrino scattering on nuclei has been proposed as a probe of electroweak physics almost 50 years ago~\cite{Freedman:1973yd} but not realized experimentally until recently~\cite{Akimov:2017ade}. Neutrinos with energies below few tens of MeV are sensitive to the entire charge of an atomic nucleus resulting in enhancement of the scattering cross sections at low energies. Due to this enhancement neutrino scattering can be probed with relatively small detectors.

The COHERENT collaboration~\cite{Akimov:2015nza} uses the Spallation Neutron Source at Oak Ridge to test coherent elastic neutrino-nucleus scattering (\cs) on several nuclei. So far, results from the CsI[Na] and Ar targets have been reported in~\cite{Akimov:2017ade,Akimov:2019rhz}, but Ge and NaI[Tl] targets are planned for the future. The main goal of the experiment is verifying the $N^2$ dependence of the cross section on the neutron number, but searching for non-standard interactions is equally interesting. Several works tackled bounds on different models utilizing \cs~\cite{Abdullah:2018ykz,Farzan:2018gtr,Denton:2018xmq,Cadeddu:2018dux,Dutta:2019eml,Cadeddu:2019eta,Khan:2019cvi,Giunti:2019xpr,Coloma:2019mbs,Han:2019zkz,Miranda:2020zji} while \cite{AristizabalSierra:2018eqm,Altmannshofer:2018xyo} concentrated on operator analysis, and \cite{Hoferichter:2020osn} provides thorough EFT analysis of \cs\ and of the relevant nuclear matrix elements.

We examine the implications of the current and future COHERENT results on the set of precision electroweak data. This is a subset of dimension 6 operators in the Standard Model (SM)~\cite{Buchmuller:1985jz,Grzadkowski:2010es} that are particularly well constrained by the LEP data as well as the measurements of the $W$-boson mass. Indirectly, because of the radiative corrections, the top quark and Higgs boson masses are important too because they contribute to the SM predictions for the relevant processes.

Model-independent operator analysis of possible deviations from the SM is by now very well established. The best known example are the $S$ and $T$ parameters~\cite{Peskin:1990zt,Peskin:1991sw} that parameterize the neutral gauge boson kinetic mixing terms and violations of the custodial symmetry, respectively. The set of tightly bounded operators is much larger than just the two corresponding to the  $S$ and $T$ parameters~\cite{Barbieri:2004qk,Han:2004az} with most constraints still dominated by the LEP experiments. In recent years, a lot of work has been devoted to operator analysis of the SM~\cite{Jenkins:2013zja,Jenkins:2013wua,Alonso:2013hga}, counting of operators~\cite{Henning:2015alf}, and constraints on the operator coefficients~\cite{Berthier:2015oma,Falkowski:2015krw,Falkowski:2017pss,Ellis:2018gqa}. The operator approach is often refereed to as the SM Effective Field Theory or SMEFT, see \cite{Brivio:2017vri} and references within. 

Computing cross sections for the \cs\ requires evaluating matrix elements of hadronic currents for the nuclei of interest. The technology of decomposing the currents into reduced matrix elements of current components with well defined spin and isospin has been established in the nuclear physics literature~\cite{DeForest:1966ycn,Donnelly:1975ze,Donnelly:1978tz}. The motivation for these developments was the study of weak interactions in nuclear processes. In \cs, the dominant spin-independent matrix elements are exact due to current conservation, but the sub-dominant matrix elements need to be computed using various applicable nuclear models. Such calculations have some degree of uncertainty, but such uncertainties do not play a large role in our result. 

This article is organized as follows. In the next section, we discuss all the ingredients of our analysis. We first enumerate the subset of precision electroweak observables that can be probed through \cs. We follow with a brief review of nuclear physics methods and matrix elements that are needed to evaluate the hadronic portion of the neutrino-nucleus scattering. We then describe the calculation of cross sections and list the experimental assumptions about the future dataset of COHERENT. In Section~\ref{SecResults}, we illustrate the bounds on the precision observables that can be obtained with the future full data set  and compare these bounds with the existing bounds obtained from other experiments. Finally, we conclude in Section~\ref{SecConclusions}.

\section{Setup and calculations }\label{SecCalc}

\subsection{Operators}
We assume that the SM Lagrangian is amended by higher-dimensional operators 
\eqn{
 \mathcal{L}=\mathcal{L}_{\rm SM} + \sum_i a_i \mathcal{O}_i, }[eq:Lagrangian]
where the sum over the operators $ \mathcal{O}_i$ and their coefficients $a_i$ is restricted to operators of interest for \cs. We consider operators of dimension six that interfere at tree level with the SM cross sections for \cs.  If interference terms are absent then such contributions are equivalent to single insertions of operators of dimension eight and an analysis restricted to operators of dimension six may not be self-consistent. We assume flavor conservation in both the lepton and quark sectors, that is consider operators with the $U(3)^5$ flavor symmetry, and also assume CP conservation. 

The following operators of dimension six appear in our analysis
\begin{eqnarray}
& \mathcal{O}_{lq}^s=\bar{l} \gamma^\mu l \,  \bar{q}\gamma_\mu q,\ \  \mathcal{O}_{lq}^t=\bar{l} \sigma^a  \gamma^\mu  l \,  \bar{q} \sigma^a \gamma_\mu  q,
      \ \ \mathcal{O}_{lu}=\bar{l} \gamma^\mu l \,  \bar{u}\gamma_\mu u, \ \ \mathcal{O}_{ld}=\bar{l} \gamma^\mu l \,  \bar{d}\gamma_\mu d, \label{eq:op4f} \\ 
 &     \mathcal{O}_{hl}^s=i (h^\dagger \overleftrightarrow{D}^\mu h)\,   \bar{l} \gamma_\mu l,
         \ \  \mathcal{O}_{hl}^t= i (h^\dagger \sigma^a \overleftrightarrow{D}^\mu h) \, \bar{l} \sigma^a \gamma_\mu l, 
         \ \ \mathcal{O}_{hq}^s=i (h^\dagger \overleftrightarrow{D}^\mu h)\,   \bar{q} \gamma_\mu q,\label{eq:opcurrent1} \\
&         \mathcal{O}_{hq}^t= i (h^\dagger \sigma^a \overleftrightarrow{D}^\mu h) \, \bar{q} \sigma^a \gamma^\mu q, 
         \ \ \mathcal{O}_{hu}=i (h^\dagger \overleftrightarrow{D}^\mu h)\,   \bar{u} \gamma_\mu u, 
         \ \ \mathcal{O}_{hd}=i (h^\dagger \overleftrightarrow{D}^\mu h)\,   \bar{d} \gamma_\mu d, \label{eq:opcurrent2} \\
 &        \mathcal{O}_S= h^\dagger \sigma^a h \, W^a_{\mu \nu} B^{\mu \nu}, \ \  \mathcal{O}_T= \left| h^\dagger D_\mu h\right|^2, \ \  \mathcal{O}_{ll}^t=\frac{1}{2}\bar{l} \sigma^a  \gamma^\mu  l \,  \bar{l} \sigma^a \gamma_\mu  l, \label{eq:SandT}
\end{eqnarray}
where $q,u,d,l,h$ denote the left-handed quarks, the right-handed up and down quarks, the left-handed leptons, and the Higgs doublet, respectively. The covariant derivative acts on the nearest field only, and $\overleftrightarrow{D}_\mu = D_\mu - \overleftarrow{D}_\mu$, while $\sigma^a$ are the Pauli matrices that act on the $SU(2)_L$ indices. Due to the assumed flavor symmetry, family indices are implicitly summed over for each type of field. There are four classes of operators listed above. First, four-fermion operators in \eqref{eq:op4f}. Second, operators that modify currents when the Higgs vacuum expectation value (vev) is substituted for $h$ in \eqref{eq:opcurrent1} and \eqref{eq:opcurrent2}. Third, the operators that correspond to the $S$ and $T$ parameters in \eqref{eq:SandT}. Fourth, $\mathcal{O}_{ll}^t$  in \eqref{eq:SandT} which does not contribute directly to \cs. However, both $\mathcal{O}_{ll}^t$ and $\mathcal{O}_{hl}^t$ contribute to the muon decay width and therefore affect determination of the Higgs vev from the Fermi coupling. 
We do not consider operators with right-handed neutrino currents, should neutrinos have Dirac masses, because such operators are very poorly constrained by COHERENT. This is because the neutrino beam in the experiment cannot contain significant fractions of right-handed neutrinos. The beams are generated from pion and muon decays that is by the charged currents, which cannot have sizable modifications. Due to the smallness of the neutrino mass, the probability of a chirality flip between neutrino production and scattering is negligible as well. This means that processes involving right-handed neutrinos are doubly suppressed: by the higher dimensional operators at both the production and detection points. 

There are two additional operators of dimension 6 that can be probed by COHERENT. These are 
\eqn{\mathcal{O}_{\nu B}=\bar{l} \, \tilde{h} \sigma^{\mu\nu} \nu_R B_{\mu \nu} + {\rm H.c.} \ \ \  {\rm and} \ \ \
    \mathcal{O}_{\nu W}=\bar{l} \sigma^a \tilde{h} \sigma^{\mu\nu} \nu_R W_{\mu \nu}^a + {\rm H.c. },}[eq:opmagnetic]
where H.c.\ denotes the Hermitian conjugate and $\tilde{h}=i \sigma^2 h^*$. A linear combination of these operators leads to the neutrino magnetic moment corresponding at low energies to the operator $\bar{\nu}_L \sigma^{\mu \nu}\nu_R  F_{\mu\nu}$, where $F_{\mu\nu}$ is the electromagnetic field strength. 
Naturalness arguments suggest that since the magnetic moment involves fields of different chirality it is proportional to the neutrino mass unless there is large tuning. The magnetic dipole moment vanishes for a single Majorana neutrino, but could exist in flavor off-diagonal form. The neutrinos could have other electromagnetic  interactions, for a review see \cite{Giunti:2014ixa} as well as other non-standard interactions that can be probed in oscillation experiments~\cite{Falkowski:2019kfn,Falkowski:2021bkq}. The bounds on the magnetic moment have been studied in~\cite{Kosmas:2015sqa,Sinev:2020bux}, so we do not include such an analysis here. The methods are however completely analogous to those underlined in the remainder of this section. Let us stress here that the assumption that chirality-changing operators are additionally suppressed by the Yukawa couplings, and therefore neglected here, is a restriction on the classes of models that one might constrain. However, large classes of models satisfy this premise, for example models of minimal flavor violation \cite{Chivukula:1987py,DAmbrosio:2002vsn}.

\subsection{Nuclear matrix elements} \label{sec:nuclear}

We now turn to the evaluation of the scattering cross section. We adopt the notation and setup in \cite{Donnelly:1978tz}. Schematically, the interaction Hamiltonian is proportional to 
\eqn{\mathcal{H} \propto j_\mu^{leptonic} {\cal J}^\mu_{hadronic} .}[eq:Hamiltonian]
The details depend on whether the interaction between the leptonic and hadronic currents is contact, as is the case of operators in \eqref{eq:op4f}, is mediated by the $Z$ boson, or is mediated by the photon. If the interaction is contact then the Hamiltonian is simply the product of the currents with the appropriate coefficient. If $Z$ mediates the interaction its propagator can be expanded in inverse powers of $m_Z^2$, and given the small momentum transfer only the leading term is kept. In case of electromagnetic interactions, which mediate interactions with the neutrino magnetic moment,  the photon propagator needs to be included in the amplitude for the process. 

Irrespectively of the type of interaction under consideration, the Hamiltonian \eqref{eq:Hamiltonian} needs to be evaluated between the initial and final states. The leptonic part is evaluated through standard perturbative methods, while the matrix elements of the hadronic current need to be evaluated for the nuclei of interest. In the case of \cs\ the initial and final nuclear states are the same, except for negligible momentum transfer. 

The isospin symmetry is broken at only a few percent level by the up-down quark mass difference and the electromagnetic interaction thus it is useful to decompose the hadronic current into the eigenstates of isospin. Since we are dealing with elastic scattering and therefore no charge transfer, the hadronic current can appear in only two isospin states with ${\cal M}_{\cal I}=0$ and ${\cal I}=0,1$, where we use the calligraphic font for the isospin and its third component. Denoting the isospin eigenstates of the current by $(J_\mu)_{\cal I M_I} $ we have 
\eqn{{\cal J}_\mu^{hadronic} = \beta_V^{(0)} (J_\mu)_{00}+  \beta_V^{(1)} (J_\mu)_{10}+\beta_A^{(0)} (J_\mu^5)_{00}+  \beta_A^{(1)} (J_\mu^5)_{10},}[eq:current]
where we further split the current into the vector and axial pieces and $\beta_{V,A}^{(0,1)}$ are numerical coefficients. Of course, for the electromagnetic current the axial pieces vanish. In terms of the quark fields, we have
\eqn{\begin{gathered} ({J}_\mu)_{00}=\frac{1}{6}\{\bar{u}\gamma_\mu u+\bar{d}\gamma_\mu d\}, \ \ 
({J}_\mu^5)_{00}=\frac{1}{2}\{\bar{u}\gamma_\mu\gamma_5 u+\bar{d}\gamma_\mu\gamma_5 d\},\\
({J}_\mu)_{10}=\frac{1}{2}\{\bar{u}\gamma_\mu u-\bar{d}\gamma_\mu d\},\ \ 
({J}_\mu^5)_{10}=\frac{1}{2}\{\bar{u}\gamma_\mu\gamma_5 u-\bar{d}\gamma_\mu\gamma_5 d\}. \end{gathered}}[eq:quarkcurrent]

Three steps are needed to get to the standard forms for the nuclear matrix elements. One uses the multipole expansion after dividing the currents into their scalar and vector parts under rotations and the resulting matrix elements are reduced using the Wigner-Eckart theorem in both the angular momentum and isospin spaces. 
Let us turn to the multipole expansion first. The currents are split into the scalar and vector parts: $J_\mu=(J_0, \vec{J})$ and the same for the axial counterpart. We call $\kappa =|\vec{q}|$ the magnitude of the three-momentum tensor. The four components of the vector current can be expanded into the following four multipoles 
\eqn{\begin{split}
  M_{{\cal J}{{\cal M}_{\cal J}};{\cal I} {\cal M}_{\cal I}}(\kappa)& =\int d^3\boldsymbol{x}\, M_{\cal J}^{{\cal M}_{\cal J}}(\kappa\boldsymbol{x}) \, {J}_0(\boldsymbol{x})_{{\cal I}{\cal M}_{\cal I}}, \ \ \ {\cal J}\ge0 \\
  {L}_{{\cal J}{{\cal M}_{\cal J}};{\cal I} {\cal M}_{\cal I}}(\kappa)& =\int d^3\boldsymbol{x}\, (\frac{i}{\kappa}\nabla M_{\cal J}^{{\cal M}_{\cal J}}(\kappa\boldsymbol{x}))\cdot \vec{{J}}(\boldsymbol{x})_{{\cal I}{\cal M}_{\cal I}}, \ \ \ {\cal J}\ge0 \\
  {T}^{el}_{{\cal J}{{\cal M}_{\cal J}};{\cal I} {\cal M}_{\cal I}}(\kappa)& =\int d^3 \boldsymbol{x}\, (\frac{1}{\kappa}\nabla\times \boldsymbol{M}_{{\cal J}{\cal J}}^{{\cal M}_{\cal J}}(\kappa\boldsymbol{x}))\cdot\vec {{J}}(\boldsymbol{x})_{{\cal I}{\cal M}_{\cal I}}, \ \ \ {\cal J}\ge1\\
   {T}^{mag}_{{\cal J}{{\cal M}_{\cal J}};{\cal I} {\cal M}_{\cal I}}(\kappa)& =\int d^3 \boldsymbol{x} \, \boldsymbol{M}_{{\cal J}{\cal J}}^{{\cal M}_{\cal J}}(\kappa\boldsymbol{x})\cdot\vec{{J}}(\boldsymbol{x})_{{\cal I}{\cal M}_{\cal I}}, \ \ \ {\cal J}\ge1 \end{split}
}[eq:multipole]
where $M_{\cal J}^{{\cal M}_{\cal J}}$ and $\boldsymbol{M}_{{\cal J}{\cal J}}^{{\cal M}_{\cal J}}$ are related to the spherical harmonics and the vector spherical harmonic, respectively, through the spherical Bessel functions of the first kind, $j_{\cal J}$, as follows 
\eqn{ M_{\cal J}^{{\cal M}_{\cal J}}(\kappa\boldsymbol{x})= j_{\cal J} (\kappa x) Y^{\cal M}_{\cal J}(\Omega_x) \ \ {\rm and} \ \ 
   \boldsymbol{M}_{{\cal J}{\cal L}}^{{\cal M}}=j_{\cal L} (\kappa x) \boldsymbol{\cal Y}^{\cal M}_{{\cal J\, L}\, 1}(\Omega_x).
}[eq:harmonics]
The multipoles in \eqref{eq:multipole} are called the Coulomb, longitudinal, transverse electric, and transverse magnetic, respectively. All these multipoles have parity $(-1)^{\cal J}$. Current conservation implies that the longitudinal matrix elements $ {L}_{{\cal J}{{\cal M}_{\cal J}};{\cal I} {\cal M}_{\cal I}}$ are proportional to $ M_{{\cal J}{{\cal M}_{\cal J}};{\cal I} {\cal M}_{\cal I}}$ and therefore not independent. 
 
Completely analogous decomposition can be made for the axial current 
\eqn{\begin{split}
  M^5_{{\cal J}{{\cal M}_{\cal J}};{\cal I} {\cal M}_{\cal I}}(\kappa)& =\int d^3\boldsymbol{x}\, M_{\cal J}^{{\cal M}_{\cal J}}(\kappa\boldsymbol{x}) \, 
   {J}_0^5(\boldsymbol{x})_{{\cal I}{\cal M}_{\cal I}}, \ \ \ {\cal J}\ge0 \\
  {L}^5_{{\cal J}{{\cal M}_{\cal J}};{\cal I} {\cal M}_{\cal I}}(\kappa)& =\int d^3\boldsymbol{x}\, (\frac{i}{\kappa}\nabla M_{\cal J}^{{\cal M}_{\cal J}}(\kappa\boldsymbol{x}))\cdot \vec{{J}}^{\, 5}(\boldsymbol{x})_{{\cal I}{\cal M}_{\cal I}}, \ \ \ {\cal J}\ge0 \\
  {T}^{el_5}_{{\cal J}{{\cal M}_{\cal J}};{\cal I} {\cal M}_{\cal I}}(\kappa)& =\int d^3 \boldsymbol{x}\, (\frac{1}{\kappa}\nabla\times \boldsymbol{M}_{{\cal J}{\cal J}}^{{\cal M}_{\cal J}}(\kappa\boldsymbol{x}))\cdot\vec {{J}}^{\, 5}(\boldsymbol{x})_{{\cal I}{\cal M}_{\cal I}}, \ \ \ {\cal J}\ge1\\
   {T}^{mag_5}_{{\cal J}{{\cal M}_{\cal J}};{\cal I} {\cal M}_{\cal I}}(\kappa)& =\int d^3 \boldsymbol{x} \, \boldsymbol{M}_{{\cal J}{\cal J}}^{{\cal M}_{\cal J}}(\kappa\boldsymbol{x})\cdot\vec{{J}}^{\, 5}(\boldsymbol{x})_{{\cal I}{\cal M}_{\cal I}}, \ \ \ {\cal J}\ge1 \end{split}
}[eq:multipole5]
where the parity of all these multipoles is $(-1)^{{\cal J}+1}$. 

Since the hadronic currents are isospin eigenstates we can write 
 \eqn{\langle {\cal I}_f {\cal M_I}_f | {\cal T}_{\cal I M_I} |  {\cal I}_i {\cal M_I}_i \rangle 
     = (-1)^{{\cal I}_f- {\cal M_I}_f} \left(\begin{array}{ccc} {\cal I}_f & {\cal I} & {\cal I}_i \\ - {\cal M_I}_f  & {\cal M_I} & {\cal M_I}_i \end{array} \right)
     \langle {\cal I}_f  \parallel  {\cal T}_{\cal I} \parallel {\cal I}_i \rangle, }[eq:W-Eisospin]
where ${\cal T}$ is any tensor that is an eigenstate of the isospin and its third component. Meanwhile, $\langle {\cal I}_f  \parallel  {\cal T}_{\cal I} \parallel {\cal I}_i \rangle$ denotes the reduced matrix element and the two by three array is the 3j symbol. 

The multipole moments have well defined angular momentum quantum numbers, so one can use the Wigner-Eckart theorem again, leading to reduced matrix elements in both spin and isospin
 \eqn{\begin{gathered}
 \langle {\cal J}_f {\cal M}_f ; {\cal I}_f {\cal M_I}_f | {\cal T}_{\cal J M ; I M_I} | {\cal J}_i {\cal M}_i ; {\cal I}_i {\cal M_I}_i \rangle 
     = (-1)^{{\cal J}_f- {\cal M}_f} \left(\begin{array}{ccc} {\cal I}_f & {\cal I} & {\cal I}_i \\ - {\cal M_I}_f  & {\cal M_I} & {\cal M_I}_i \end{array} \right)
        \\ \times 
        (-1)^{{\cal I}_f- {\cal M_I}_f} \left(\begin{array}{ccc} {\cal I}_f & {\cal I} & {\cal I}_i \\ - {\cal M_I}_f  & {\cal M_I} & {\cal M_I}_i \end{array} \right)
     \langle {\cal J}_f;  {\cal I}_f  \parallel  {\cal T}_{\cal J; I} \parallel {\cal J}_i ; {\cal I}_i \rangle, \end{gathered} }[eq:W-Eidouble]
where now $ \langle {\cal J}_f;  {\cal I}_f  \parallel  {\cal T}_{\cal J; I} \parallel {\cal J}_i ; {\cal I}_i \rangle$ denotes the twice reduced matrix element. We do not introduce different symbols for the twice reduced matrix elements as the quantum numbers of the operator make it clear which reduction(s) took place.

It is clear that the multipole expansion is in powers of $(\kappa R)^{\cal J}$, where $R$ is a typical nucleus radius and $\kappa$ is the momentum transfer.  A good estimate is  $1/R=Q\approx 250~{\rm MeV}$ that is the typical momentum of nucleons in nuclei. Given that for the neutrinos detected by COHERENT the magnitude of the three-momentum transfer $\kappa$ is small compared to $Q$ we can concentrate on the lowest non-vanishing multipoles only. Due to their negative parity, the matrix elements of the axial current with ${\cal J}=0$, and in general with even ${\cal J}$, vanish in elastic scattering. An analysis of the low-energy limit shows that the leading matrix elements are those of $M_{0;0}$, $M_{0;1}$, $L^5_{1,0}$, and $L^5_{1,1}$~\cite{Donnelly:1978tz}. (Of the same order are also matrix elements of ${T}^{el_5}_{1,0}$ and ${T}^{el_5}_{1,1}$, but these are related to $L^5_{1,0}$, and $L^5_{1,1}$.)

The Coulomb matrix elements are computed easily since they are related to conserved charges
\eqn{ \begin{split}
  \langle J;T \parallel  M_{0; 0} \parallel J;T\rangle &= \frac{1}{2 \sqrt{4\pi}} A \sqrt{2J+1}\sqrt{2T+1}, \\
  \langle J;T \parallel  M_{ 0; 1} \parallel J;T\rangle  &= \frac{1}{\sqrt{4\pi}}  \sqrt{T(T+1)}\sqrt{2J+1}\sqrt{2T+1},
  \end{split}
}[eq:Coulomb]
where $A$ is the atomic number, $J$ spin of the nucleus, and $T$ its isospin. In terms of the number of neutrons and protons,  respectively $N$ and $Z$, $A=N+Z$ and $T=\frac{1}{2}|Z-N|$.

The matrix elements of $L^5_{1,0}$ and $L^5_{1,1}$ vanish for nuclei with no spin since these operators carry non-zero angular momentum. For nuclei with spin we use results of two different calculations. The $^{23}Na$,  $^{127}I$, and $^{133}Cs$ matrix elements in Table~\ref{table:matrix} are adopted from \cite{Hoferichter:2020osn,MenendezHoferichter}, while the $^{204}Tl$ matrix elements are results by Pirinen and Ydrefors~\cite{Pirinen}. The values are listed in Table~\ref{table:matrix}. 
\begin{table}[!ht]
\centering
\begin{tabular}{|c|c|c|c|c|}
\hline
Nucleus &J&T&$  \langle J;T \parallel L^5_{1;0}\parallel J;T\rangle$ &$  \langle J;T \parallel| L^5_{1;1}\parallel J;T\rangle$\\
\hline 
$^{23}Na$ & $\frac{3}{2}$ & $\frac{1}{2}$&-0.0612&0.197\\
\hline
$^{127}I$& $\frac{5}{2}$ & $\frac{21}{2}$ &-0.346&0.698\\
\hline
$^{133}Cs$& $\frac{7}{2}$ & $\frac{23}{2}$ &0.363&-0.878 \\
\hline
$^{204}Tl$& $2$ & 21 &-0.1482&0.0056\\
\hline
\end{tabular}
\caption{\label{table:matrix} Longitudinal matrix elements from shell model calculations \cite{Hoferichter:2020osn,MenendezHoferichter} ($Na, I, Cs$)  and \cite{Pirinen} ($Tl$).}
\end{table}
Estimating error bars on these matrix elements is not straightforward. It is likely safe to assume that such errors are in the $10-30\%$ range. A comparison between model computations and experimental values of energy levels and ground state magnetic moment support this estimate~\cite{Pirinen}. Numerous works are devoted to computing the matrix elements relevant for \cs, see for example \cite{Papoulias:2019lfi,Kronfeld:2019nfb,Payne:2019wvy,Hoferichter:2020osn}.

\subsection{Cross sections}

Computing the cross sections is now straightforward. With the $Z$ propagator truncated to the momentum-independent part, the interaction Hamiltonian is
\eqn{\begin{split}
   \mathcal{H} &=\frac{G_F}{\sqrt{2}} j_\mu^{leptonic} {\cal J}^\mu_{hadronic} \\
   &=\frac{G}{\sqrt{2}}\sum\limits_{q=u,d}\left[\bar{\nu}\gamma_\mu(1-\gamma_5)\nu\right] \, \left[(f^{qL}+\epsilon^{qL})(\bar{q}\gamma^\mu(1-\gamma_5)q)+(f^{qR}+\epsilon^{qR})(\bar{q}\gamma^\mu(1+\gamma_5)q)\right],
    \end{split}  }[eq:H]
where $G_F$ is the Fermi constant. The couplings $f^{qL,qR}$ are the SM couplings, while $\epsilon^{qL,qR}$ are the deviations from the SM values due to the higher-dimensional operators. These couplings are 
\eqn{ \begin{gathered}
 f^{uL}=\frac{1}{2}-\frac{2}{3}\sin^2\theta_W, \ \ \ f^{dL}=-\frac{1}{2}+\frac{1}{3}\sin^2\theta_W, \\
f^{uR}=-\frac{2}{3}\sin^2\theta_W, \ \ \ f^{dR}=\frac{1}{3}\sin^2\theta_W,
  \end{gathered}}[eq:coupligsf]
where $\theta_W$ is the Weinberg angle. Meanwhile, the $\epsilon$'s are given by 
\eqn{ \begin{split}
\epsilon^{uL}&=-\frac{v^2}{2}(a^s_{lq}+a^t_{lq}+a_{hq}^s-a_{hq}^t +  f^{uL} \Delta_1 + \frac{2}{3} \Delta_2), \\
\epsilon^{dL}&=-\frac{v^2}{2}(a^s_{lq}-a^t_{lq}+ a_{hq}^s+ a_{hq}^t+f^{dL} \Delta_1-\frac{1}{3} \Delta_2), \\
\epsilon^{uR}&=-\frac{v^2}{2}(a_{lu}+  a_{hu}+f^{uR} \Delta_1+\frac{2}{3} \Delta_2), \\
\epsilon^{dR}&=-\frac{v^2}{2}(a_{ld}+ a_{hd}+f^{dR} \Delta_1-\frac{1}{3} \Delta_2), 
\end{split} }[eq:epsilons]
where $v$ is the electroweak vev and $G_F=\frac{1}{\sqrt{2}v^2}$. The contributions $\Delta_1$ and $\Delta_2$ are universal affecting all terms. $\Delta_1$ arises from modification of the $\nu$-$Z$ coupling and the additional contributions to $G_F$, while $\Delta_2$ comes from the shift in the value of the Weinberg angle caused by the operators ${\cal O}_S$ and ${\cal O}_T$ and those that contribute to $G_F$ as well~\cite{Han:2004az,Burgess:1993mg,Skiba:2010xn}. Their values are 
\eqn{ \begin{split} 
    \Delta_1&=2 (a_{hl}^s+ a_{hl}^t -  a_{ll}^t+\frac{1}{2} a_T), \\
    \Delta_2&= \tan (2 \theta_W) \left[a_S + \frac{ \sin (2 \theta_W)}{2} \left(2 a_{hl}^t -  a_{ll}^t + \frac{1}{2} a_T \right)\right].
    \end{split}   }[eq:Deltas]
    
The differential scattering cross section is given in \cite{Donnelly:1978tz} in terms of the coefficients $\beta^{(0,1)}_{V,A}$ introduced in \eqref{eq:current} and the reduced  matrix elements introduced in Section~\ref{sec:nuclear}. In the limit of vanishing momentum transfer $\boldsymbol{q}^2$ 
\eqn{ \begin{split}
  \left. \frac{d\sigma}{dE}\right|_{\boldsymbol{q}^2 \rightarrow 0}&=\frac{4G_F^2M}{(2J+1)(2T+1)}\left[
                \phantom{\left|\beta_V^{(0)}\langle J;T\parallel {M}_{0;0} \parallel J;T\rangle
  +\frac{M_T}{\sqrt{T(T+1)}}\beta_V^{(1)}\langle J;T\parallel {M}_{0;1} \parallel J;T\rangle\right|^2} \right. \\
  & \left(1-\frac{ME}{2E_{\nu}^2}\right)\left|\beta_V^{(0)}\langle J;T\parallel M_{0;0}\parallel J;T\rangle
  +\frac{M_T}{\sqrt{T(T+1)}}\beta_V^{(1)}\langle J;T\parallel M_{0;1}\parallel J;T\rangle \right|^2  \\
  &  \left. + \left(1+\frac{ME}{2E_{\nu}^2}\right)
     \left|\beta_A^{(0)}  \langle J;T\parallel L^5_{1;0}\parallel J;T \rangle +\frac{M_T}{\sqrt{T(T+1)}}\beta_A^{(1)}  \langle J;T\parallel L^5_{1;1}\parallel J;T \rangle  
           \right|^2\right],
 \end{split} }[eq:cross-section]
where $M$ is the nucleus mass, $E_\nu$ the energy of the incoming neutrino, and $M_T=\frac{1}{2}(Z-N)$ the third component of the isospin. Comparing \eqref{eq:current} and \eqref{eq:quarkcurrent} with \eqref{eq:H} it is straightforward to obtain 
\eqn{ \begin{gathered}
 \beta_V^{(0)}=-2\sin^2\theta_W+3(\epsilon^{uL}+\epsilon^{uR}+\epsilon^{dL}+\epsilon^{dR}),
    \ \ \  \beta_A^{(0)}=-\epsilon^{uL}+\epsilon^{uR}-\epsilon^{dL}+\epsilon^{dR}, \\
  \beta_V^{(1)}=1-2\sin^2\theta_W+\epsilon^{uL}+\epsilon^{uR}-\epsilon^{dL}-\epsilon^{dR}, \ \ \ \beta_A^{(1)}=-1-\epsilon^{uL}+\epsilon^{uR}+\epsilon^{dL}-\epsilon^{dR}.
\end{gathered} }[eq:betas]
 It is clear that COHERENT is sensitive to four linear combinations of the coefficients, the ones appearing above in \eqref{eq:betas}. We will come back to this point later on. 

\subsection{Detectors and the neutrino beam}

The COHERENT experiment is going to use four different detectors. So far, results for only two of these four have been reported~\cite{Akimov:2017ade,Akimov:2019rhz}. To determine the future sensitivity of the experiment we assume the detector parameters as in \cite{Barbeau}. An energy-averaged detection efficiency of 50\% is assumed for each detector.
\begin{table}[ht]
\centering
\begin{tabular}{|c|c|c|c|c|}
\hline
Nuclear Target&Mass [kg]&\tabincell{c}{Distance from\\source [m]} & \tabincell{c}{Recoil threshold\\\protect{[keVr]}} &\tabincell{c}{Quenching \\factor}\\
\hline 
CsI[Na] & 14 & 20&6.5&7\%\\
\hline
Ge& 10 & 22 &5&2\%\\
\hline
LAr&35&29&20&25\%\\
\hline
NaI[Tl]&2000&22&13&15\%\\
\hline
\end{tabular}
\caption{\label{tab:detectors} Parameters used in the calculation for the four detectors~\cite{Barbeau,Akimov:2015nza}. There is a 10\% uncertainty in neutrino flux aside from the uncertainties listed in the table.}
\end{table}
The elemental composition of the CsI[Na] and NaI[Tl] detectors is displayed in Table~\ref{table:composition}.
\begin{table}[!h]
\centering
\begin{tabular}{|c|c|c|c|c|c|}
\hline
Element&Atomic weight&Mass percentage\\
\hline 
Cs & 133 & 47\%\\
\hline
I& 127 & 45\%\\
\hline
Na&23&8\%\\
\hline
\end{tabular}
\hfill
\begin{tabular}{|c|c|c|c|c|c|}
\hline
Element&Atomic weight&Mass percentage\\
\hline 
Na&23& 6.5\%\\
\hline
I& 127 & 35.8\%\\
\hline
Tl&204&57.7\%\\
\hline
\end{tabular}
\caption{\label{table:composition}Mass percentage of each element in CsI[Na] and NaI[Tl].}
\end{table}

Given the inputs in tables \ref{tab:detectors} and \ref{table:composition}, the total number of events in a detector is calculated as
\eqn{
N_{events}=t \phi\frac{M_{detector}}{M}\int\limits_{E_{min}}^{E_{max}}dE_{\nu}\int\limits_{E_{th}}^{E_{recoil~ max}}dE\lambda(E_{\nu})\frac{d\sigma}{dE}(E_{\nu},E),
}[eq:events]
where $t$ is the data taking time period and  $\phi$ is the neutrino flux. In this analysis, we use the following expression to obtain the product of $t$ and $\phi$: $t\phi=rN_{POT}/4\pi L^2$~\cite{Cadeddu:2017etk}, where $r= 0.08$ is the number of neutrinos per flavor that are produced for each proton on target, $N_{POT}=1.76\times10^{23}$ is the number of proton on target for a live time $t\sim1$~year~\cite{Akimov:2017ade} and $L$ is the distance between the source and the COHERENT detector. Furthermore, $\lambda(E_{\nu})$ is the normalized neutrino spectrum that is the sum of the $\nu_{e}$ and $\bar{\nu}_\mu$ spectra from the $\mu^+$ decays
\eqn{ \begin{split}
\lambda_{\nu_e}&=\frac{96}{m_\mu^3}E_{\nu}^2(1-\frac{2E_{\nu}}{m_\mu}), \\
\lambda_{\bar{\nu}_\mu}&=\frac{48}{m_\mu^3}E_{\nu}^2(1-\frac{4E_{\nu}}{3m_\mu}),
 \end{split} }[eq:spectrumdelayed]
with the maximum energy of $E_{max}=m_{\mu}/2$, $m_\mu$=105.6~MeV, and the mono-energetic $\nu_{\mu}$'s from the $\pi^+$ decay
\eqn{
\lambda_{\nu_\mu}=\delta\left(E_\nu-\frac{m_\pi^2-m_\mu^2}{2 m_\pi}\right).
}[eq:spectrumprompt]
The minimum incoming neutrino energy required for detection is determined by the detector's threshold energy $E_{th}$ and the nucleus mass $M$ through the relation $E_{th}=2E_{min}^2/(M+2E_{min})$. 

\section{Results}\label{SecResults}

We compute the number of events for each of the detectors as a function of the coefficients $a_i$ in \eqref{eq:Lagrangian}. The number of events in each detector is combined into a $\chi^2$ distribution through 
\eqn{ \begin{split}
\chi^2_{tot}&=\sum\limits_X\frac{(N_X(a_i)-N_{X,exp})^2}{\sigma_X^2} =\frac{(N_{CsI[Na]}(a_i)-N_{CsI[Na]}^{SM})^2}{N_{CsI[Na]}^{SM}\times1.17} \\
&+\frac{(N_{Ge}(a_i)-N_{Ge}^{SM})^2}{N_{Ge}^{SM}\times1.12} +\frac{(N_{Ar}(a_i)-N_{Ar}^{SM})^2}{N_{Ar}^{SM}\times1.35}+\frac{(N_{NaI[Tl]}(a_i)-N_{NaI[Tl]}^{SM})^2}{N_{NaI[Tl]}^{SM}\times1.25},
 \end{split} }[eq:chi2]
 where in the absence of full experimental results we assumed perfect agreement with the SM. The standard deviations are estimated from the Poisson distribution and additional uncertainties as 
\eqn{ \sigma_X=\sqrt{N^{SM}\times(1+{\rm quenching\ factor+10\%~neutrino\ flux\ uncertainty})}. }[eq:sigma]
 The uncertainties of the matrix elements of the longitudinal operators are negligible in the error budget because the Coulomb matrix elements dominate.
 
Before we describe the results we want to briefly comment on the energy scales in the problem. The effective Lagrangian in \eqref{eq:Lagrangian} and the corresponding operators \eqref{eq:op4f} through \eqref{eq:SandT}  are defined at or above the Higgs mass scale. Below the electroweak scale, the $SU(3)\times SU(2) \times U(1)_Y$ invariant operators are matched into operators invariant under $SU(3)\times U(1)_{em}$, which are the four fermion-operators in \eqref{eq:H} which are in turn matched at the QCD scale to nuclear matrix elements. A complete basis of operators below the electroweak scale is described in \cite{Jenkins:2017jig} and their one-loop renormalization group evolution (RGE) equations in \cite{Jenkins:2017dyc}. The operators of interest here, which are products of neutrino current and quark currents, evolve under the RGE proportionately to  
 the electromagnetic coupling but do not have any contributions from the strong coupling. Thus, the coefficients change insignificantly between the weak and QCD scales at a few percent level. This estimate is verified by explicit numerical running of the coefficients using the code implemented in \cite{Aebischer:2018bkb}. 
 
 For the individual coefficients $a_i$ in \eqref{eq:Lagrangian}, a comparison of bounds obtained by a global fit to low-energy and collider experiments obtained in \cite{Falkowski:2017pss} and those one will be able to extract from the future COHERENT data is presented in Table~\ref{table:bounds}. In the table below as well as in the figure later in this section we have used the bounds in \cite{Falkowski:2017pss} in the flavor symmetric case.
 \begin{table}[!ht]
\centering
\begin{tabular}{|c|c|c|c|c|c|}
\hline
Coefficient & Existing bounds [${\rm GeV}^{-2}$]~\cite{Falkowski:2017pss}&COHERENT experiment [${\rm GeV}^{-2}$]\\
\hline 
\hline 
$a_{lq}^s$&$-1.2\times10^{-8}<a_{lq}^s<3.9\times10^{-8}$&$|a_{lq}^s|<2.2\times10^{-8}$\\
\hline
$a_{lq}^t$&$-0.6\times10^{-9}<a_{lq}^t<1.5\times10^{-8}$&$|a_{lq}^t|<3.7\times10^{-7}$\\
\hline
$a_{hl}^s$&$-7.0\times10^{-9}<a_{hl}^s<7.5\times10^{-9}$&$|a_{hl}^s|<1.2\times10^{-7}$\\
\hline
$a_{hl}^t$&$-8.3\times10^{-9}<a_{hl}^t<0.4\times10^{-9}$&$|a_{hl}^t|<1.3\times10^{-7}$\\
\hline
$a_{hq}^s$&$-1.7\times10^{-8}<a_{hq}^s<8.9\times10^{-9}$&$|a_{hq}^s|<2.2\times10^{-8}$\\
\hline
$a_{hq}^t$&$-8.9\times10^{-9}<a_{hq}^t<1.7\times10^{-8}$&$|a_{hq}^t|<3.7\times10^{-7}$\\
\hline
$a_{lu}$&$-1.8\times10^{-8}<a_{lu}<9.2\times10^{-8}$&$|a_{lu}|<4.7\times10^{-8}$\\
\hline
$a_{ld}$&$-4.8\times10^{-9}<a_{ld}<1.1\times10^{-7}$&$|a_{ld}|<4.2\times10^{-8}$\\
\hline
$a_{hu}$&$-2.5\times10^{-8}<a_{hu}<5.7\times10^{-8}$&$|a_{hu}|<4.7\times10^{-8}$\\
\hline
$a_{hd}$&$-1.1\times10^{-8}<a_{hd}<1.0\times10^{-7}$&$|a_{hd}|<4.2\times10^{-8}$\\
\hline
$a_{ll}^t$&$-1.2\times10^{-8}<a_{ll}^t<0.2\times10^{-9}$&$|a_{ll}^t|<3.9\times10^{-6}$\\
\hline
$a_{S}$&$-8.9\times10^{-9}<a_{S}<1.7\times10^{-9}$&$|a_{S}|<5.2\times10^{-8}$\\
\hline
$a_{T}$&$-2.2\times10^{-8}<a_{T}<2.6\times10^{-9}$&$|a_{T}|<7.8\times10^{-6}$\\
\hline
\end{tabular}
\caption{\label{table:bounds} Comparison between present limits and the ones obtained from COHERENT at 90\% C.L., taking one parameter at a time.}
\end{table}
While none of the individual bounds from COHERENT are obviously more stringent than the existing ones, two points are apparent. First, when the bounds on a coefficient are comparable between the two columns in Table~\ref{table:bounds}, for example on $a_{lq}^s$ or $a_{lu}$, combining the COHERENT data with all the other precision electroweak data will improve the bounds. Second, the bounds on the individual coefficients are not the whole story. It is the combined fit to all the coefficients together, or in other words to arbitrary linear combinations of the coefficients, that is useful in constraining new physics~\cite{Han:2004az,Skiba:2010xn}. In the space of $n$ operator coefficients it is the $n$-dimensional ellipsoid that encodes the full experimental information.

\begin{figure}[!th]
    \centering
    \begin{subfigure}[t]{0.45\textwidth}
        \centering
        \includegraphics[width=\linewidth]{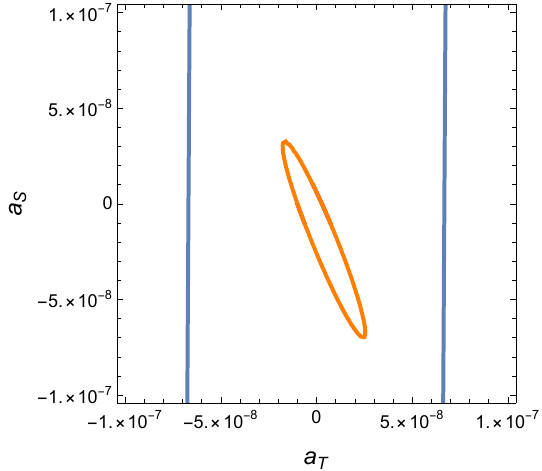} 
    \end{subfigure}
    \hfill
    \begin{subfigure}[t]{0.45\textwidth}
        \centering
        \includegraphics[width=\linewidth]{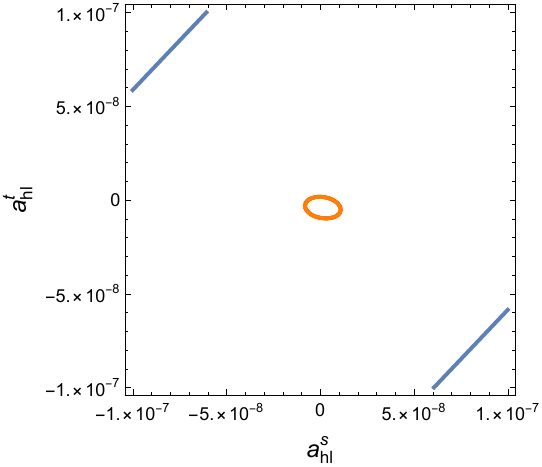} 
    \end{subfigure}

    \vspace{0.5cm}
  \centering
    \begin{subfigure}[t]{0.45\textwidth}
        \centering
        \includegraphics[width=\linewidth]{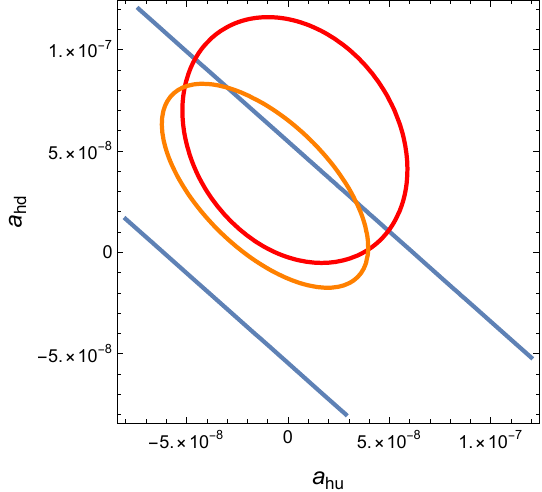} 
    \end{subfigure}
    \hfill
    \begin{subfigure}[t]{0.45\textwidth}
        \centering
        \includegraphics[width=\linewidth]{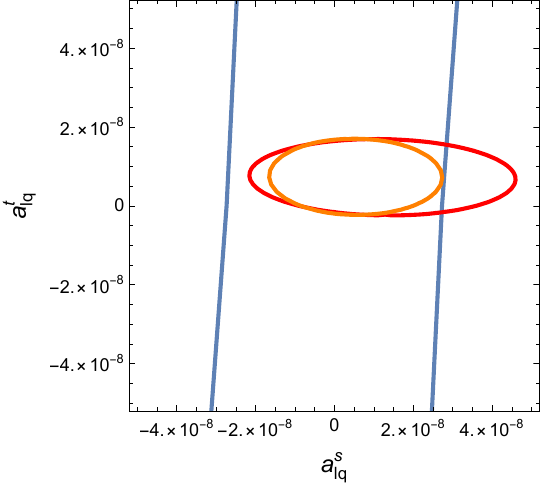} 
    \end{subfigure}

    \caption{\label{fig:bounds} Comparison between the bounds in~\cite{Falkowski:2017pss} and future COHERENT bounds projected onto planes of two coefficients at 90\% confidence levels. The COHERENT bounds are the nearly parallel lines in the plots (blue). The existing bounds are the larger ellipses (red) and the combined bounds are the inner ellipses (orange). For the plots in the top row the red and orange curves overlap showing that COHERENT will not improve bounds on the operators in those plots. }
\end{figure}

The plots in Figure~\ref{fig:bounds} exemplify the main outcome of this analysis. For some coefficients, the existing limits are so stringent that \cs\ will not deliver improvements unless the amount of data is much larger than the projected quantity. However, for certain coefficients a combined fit that includes the COHERENT data will provide improvements. The advantage of COHERENT is that it is sensitive to different directions in the space of operators, compared to other experiments, and it is in these unique directions where there will be most improvement from the full data set.

\section{Conclusions}\label{SecConclusions}

Experimental observation of \cs\ certainly opened up an interesting new regime for neutrino physics. We have examined the impact of future COHERENT dataset, consisting of data from four different detectors, on the body of precision electroweak observables. 

A demonstration of the COHERENT experiment's potential is contained in Table~\ref{table:bounds} and Figure~\ref{fig:bounds}. There, we presented future bounds on both the individual coefficients of operators and select two-dimensional projections of the $\chi^2$ function for the 13 operators considered in this article. 

It is clear that the COHERENT results will need to be eventually included in the complete fit of all precision electroweak data. For some of the operators, the ones with existing stringent bounds, one cannot expect any improvement. There are some operators, however, for which inclusion of the COHERENT dataset will yield tighter bounds. Exactly how big this improvement will be is impossible to predict exactly since it will depend on how well the data will agree with the SM\@. Potential deviations, even if purely statistical in nature, will affect the full fit. 

The COHERENT data is sensitive to four linear combinations of coefficients of operators. These are listed in \eqref{eq:betas} in terns of parameters $\epsilon$ introduced in ~\eqref{eq:epsilons}. However, for any particular nucleus there are two linear combinations of coefficients that enter the cross section formula in \eqref{eq:cross-section}. Of course, since deviations from the SM are obtained from the interference terms between higher-dimensional operators and SM processes there is actually only one linear combination that can be teased out with one measurement. This is true even with detectors that contain several nuclei. Therefore, with one detector only a single  direction in the space of operators can be bounded by \cs. Nevertheless, with several detectors all four combinations can be bounded independently as variations in nuclear matrix elements among nuclei pick different admixtures of the four underlying combinations in \eqref{eq:betas}. A potential caveat is that there will be different amounts of data from different detectors, so not every one of the four combinations will be equally well constrained. If improving the bounds on the four combinations in \eqref{eq:betas} were a priority one would need to rethink the balance between the amount of data taken with different detectors to maximize the potential for obtaining independent constraints. 


\ack{
The authors would like to thank Y.~Alhassid and F.~Iachello for helpful conversations, J.~Suhonen for correspondence, and P.~Pirinen for providing us with the results of calculations of nuclear matrix elements.  We are also grateful to A.~Khan for correspondence, and J.~Men\'endez and M.~Hoferichter for extensive correspondence and their detailed comments on nuclear matrix elements. This work was supported in part by DOE HEP grant DE-SC00-17660.}

\newpage

\bibliography{Bibliography}


\end{document}